# Magneto-transport and first principle study of strong topological insulator gray-Arsenic

N. K. Karn[1,2], Kapil Kumar[1,2], Geet Awana[3], Kunal Yadav[3], S. Patnaik[3], V.P.S. Awana[1,2]

*[1]Academy of Scientific and Innovative Research (AcSIR), Ghaziabad 201002, India*

*[2]CSIR-National Physical Laboratory, Dr. K. S. Krishnan Marg, New Delhi-110012, India*

*[3]School of Physical Sciences, Jawaharlal Nehru University, New Delhi, India, 110067*

**Abstract:**

This article reports on the synthesis of single crystalline gray-Arsenic (As) via the Bismuth flux method. The X-ray Diffraction (XRD) pattern revealed the single phase of the as-grown crystal, which crystallized in the rhombohedral structure with the space group R-3m. The sharp XRD peaks observed on mechanically exfoliated thin flakes of the same ensured high crystallinity of the same with growth direction along the c-axis. The Energy Dispersive X-ray Analysis (EDAX) endorses the stoichiometric purity of the as-grown As single crystal. The Raman spectra are recorded to study the vibrational mode, which showed peaks at 196.2cm$^{-1}$ and 255.74cm$^{-1}$, identified as $E_g$ and $A_{1g}$ modes respectively, by DFT calculations. The as-grown crystal is further characterized for its electronic and magneto-transport properties. The resistivity vs. temperature (ρ-T) measurements illustrated its metallic nature throughout, right from 300K down to 2K. The measured residual resistivity ratio ($\rho_{300K}/\rho_{2K}$) of the sample is 180, which endorses the high metallic nature of the as-synthesized As single crystal. The transverse magnetic field-dependent resistivity (ρ-H) measurements elucidated huge ($10^4$%) magneto-resistance (MR) at 2K and 14Tesla transverse magnetic fields, along with the SdH oscillations, indicating the presence of topological surface states. The non-trivial band topology and edge states in As are confirmed by first principle calculations. Not only do orbital projected bands show the signature of band inversion but also the Z2 invariant value (1,111) calculated by Wilson's loop method affirms As to be a strong topological insulator (TI). Clear evidence of topological edge states in plane $k_z = 0$ has been observed in surface state spectra and slab bands.

*Corresponding Author
N. K. Karn: Senior Research Fellow
CSIR-NPL, New Delhi, India, 110012
E-mail: nkk15ms097@gmail.com
Ph. +91-11-45609357

## Introduction:

The enormously high magneto-resistance (MR) has become a pivotal field in fundamental and applied physics [1]. In recent years, high MR has emerged as a transport characteristic of the swiftly advancing topological materials [2]. This effect unfolds owing to strong spin-orbit coupling (SOC), observed in pnictides, selenides, and heavy element-based systems [3, 4]. In the case of lighter elements other mechanisms such as Rashbha coupling and correlation effects give rise to the SOC effect and ensuing high MR [5,6].

In the past two decades, topological materials have emerged as revolutionary materials owing to the demonstration of rich phenomenology, such as different versions of the Hall effect [7-9], the thermal Hall effect [10], wide applications in spintronics [11, 12] and topological superconductors hosting Majorana Fermions [13]. Such materials have characteristic topological surface states, leading to an efficient transport phenomenon. Methods to probe topological surface states are swiftly evolving; experimentally a direct way is angular-resolved photoemission spectroscopy (ARPES) measurements, where linear band dispersion with the Dirac cone can be observed [14]. Indirectly, these can be probed by detecting electromagnetic responses in magneto-transport measurements such as non-saturating linear MR (LMR) [15, 16] or negative MR [17] and oscillations in MR (SdH oscillations) [18]. Theoretically, topological surface states are computed and characterized by associated invariants viz. Z2 invariant [19]. There have been thousands of materials theoretically analyzed, and surprisingly 30% of them are attributed to topologically non-trivial phases [20]. Yet, materials in their elemental form show non-trivial topological features rarely. In group III elements, the topology of Bismuth is contrasting as experiments show non-trivial surface states, whereas theoretically, it is trivial [21]. Unlike Bismuth, Arsenic has not been studied widely, however, it has shown potential as a possible good TI candidate in terms of its Giant MR [22] and non-trivial topological surface states [23, 24].

Pure Arsenic, despite being a noxious element, in the last few years has gathered interest from physicists and material scientists as a potential quantum material. Being a metalloid, As has several allotropes with different transport properties: semi-metallic behavior of gray-As [22], semiconducting black-As [25], and the mixed structure of the two orthorhombic As i.e., pararsenolamprite [26]. Gray-As has a rhombohedral structure of puckered honeycomb layers, like black Phosphorus. Gray-As has a zero bandgap in bulk or multilayer form, contrary to black Phosphorous. But the same opens a bandgap of 1.2–1.4eV [4] when amorphized or reduced to bilayers or monolayers. With such features, gray-As stands as a promising candidate for tunable bandgap engineering and novel device applications.

Recently, two reports showed the typical quantum oscillations in magneto-resistivity measurements of gray-As [22, 27] with differences in their synthesis procedure. The Bismuth flux method results in crystal growth along the (0 1 2) direction [27], whereas the modified Bridgeman technique results in the (0 0 3) growth direction [22]. Both show out-of-plane anisotropy in MR. In the present work, we have used a novel Bismuth flux two-step method for producing (0 0 3) growth direction of As crystals. The vibrational modes have been studied by Raman spectroscopy and verified by density functional theory(DFT) based calculations. The out-of-plane magneto-transport measurement shows distinct SdH oscillations and the calculation of the Berry phase showed the non-trivial topological nature of gray-As single crystal. The computed angle dependent dHvA frequencies gives insights into the conduction by surface states. Further, the Z2 invariant value of gray-As obtained by Wilson's loop method implemented in first principle methods establishes the same as a strong non-trivial TI.

## Experimental and Computational Methods

The single crystal of As is grown by using the Bismuth flux by following an optimized two-step heat treatment as shown in Fig. 1. In the first step, high-quality powders Sigma of Bi (99.999%) and As (99.999%) and were taken in 1:1 stoichiometric ratio and the resultant mixture was grounded by agate mortar pestle in MBraun glove box filled with argon gas. The resulting homogeneous mixture is then converted into a cylindrical pallet using a hydraulic press. Further, it is vacuum encapsulated in a quartz ampoule at a pressure of $5\times10^{-5}$mbar. The sealed ampoule of As is heated to $500^0$C at a rate of $60^0$C/h in a simple programmable muffle furnace. The ampoule dwelt at this temperature for 48hours to make the melt homogenous and then cooled down to room temperature. In step II, the molten chunk is again powdered, palletized, and vacuum encapsulated, the same as was done before for the first heat treatment. The encapsulated pallet is then subjected to heat treatment as shown in Fig. 1(b). The obtained crystals are though gray but shiny and are embedded in the matrix of silvery shiny Bismuth. The embedded As crystals are mechanically cleaved out having a size of ~3mm. The inset of Fig. 2(a) displays the Bi matrix embedded with As crystals and mechanically cleaved As crystals.

The grown arsenic (As) crystals are analyzed for their phase purity, growth orientation, and compositional purity. For phase purity, XRD patterns are collected using a Rigaku Miniflex

II tabletop XRD instrument with Cu-Kα radiation (wavelength of 1.541(8)Å) on both crystal flakes and gently crushed powder samples. The powder XRD data are analyzed using Full Proof software. High-resolution transmission electron microscopy (HRTEM) and selected area electron diffraction (SAED) patterns are recorded using a JEOL 2100F instrument at resolutions of 1nm and $2nm^{-1}$, respectively, with data analysis performed using Image-J software. For morphology and chemical purity, SEM and EDAX measurement have been performed using a Zeiss EVO MA 10 microscope equipped with an Oxford INCA 250 system. Raman spectroscopy, performed on a Renishaw inVia Raman Reflex system with a 514nm laser, has been used to study vibrational modes. Magneto-transport measurements are carried out utilizing a Quantum Design Physical Property Measurement System (PPMS). The sample is mounted on a PPMS resistivity puck with a standard four-probe electrical contact configuration, where the current flows in the ab-plane and the magnetic field is kept transverse to the current direction.

For the theoretical analysis, we have followed the computational methodology as in refs. [28, 29]. Density Functional Theory (DFT) has been employed to compute the electronic state and band structure of the As crystal using the open-access software Quantum Espresso [30]. The calculations utilize Perdew-Burke-Ernzerhof (PBE) type pseudo-potentials, which include corrections for exchange correlations [31]. For spin-orbit coupling (SOC) calculations, standard norm-conserving pseudo-potentials with full relativistic corrections from the PSEUDODOJO library have been used. The k-space has been sampled on an 8×8×8 Monkhorst-Pack mesh [32]. The following parameters have been set for the calculations: charge cutoff at 336Ry, wfc cutoff at 84Ry, degauss at 0.02, and SCF electronic convergence cutoff at $10^{-10}$Ry. Additionally, out of 36 SOC Bloch bands, 16 bands have been wannierized with projections in s and p orbitals, using a tolerance of $10^{-11}$ in Wannier90 [33], and Wannier charge centers (WCC) have been calculated. To determine topological properties, the tight-binding Hamiltonian generated by Wannier90 has been implemented in Wannier Tools [34], and the Z2 invariant has been computed. The iterative surface Green's function method [35,36] has been employed to compute the surface state spectral function along the (001) plane, as implemented in Wannier Tools. Vibrational modes have also been calculated without the SOC parameter, using norm-conserving pseudo-potentials.

## Results and Discussion

Gray-As has a rhombohedral lattice structure with centro-symmetric space group R-3m which is iso-structural with Bismuth. After following the two-step heat treatment shiny gray-As crystals are obtained by mechanical cleaving. Fig. 2(a) shows the single crystal XRD peaks of cleaved as-grown gray-As crystal. The crystal planes are aligned along (0 0 3n) direction indicating that the growth direction is along the c-axis. One tiny peak is observable just after the (0 0 3) plane peak, hinting towards the presence of a tiny impurity. The inset of Fig. 2(a) shows the matrix of As-embedded Bi and mechanically cleaved As crystal pieces from the matrix. From single crystal XRD, the calculated lattice parameter c is 10.546(5)Å which is comparable to previous reports [22] and in agreement with PXRD Rietveld refinement results. Fig. 2(b) shows the powder XRD of As crystals and its Rietveld refinement in the rhombohedral phase. All the observed peaks are well-fitted and thereby obtained refined lattice parameters are shown in Table 1. The $\chi^2$ value is the measure of goodness of fit, which is found to be 2.22 within the acceptable range. Using the refined parameters, the rhombohedral unit cell of gray-As is drawn by VESTA software as shown in the inset of Fig. 2(b). The Wykoff position of As is (0, 0, 0.22874). The impurity peaks are observed in the PXRD of gray-As and are marked with (*), which belongs to Bismuth. This is due to leftovers of Bi present in the powder during the mechanical separation process of As crystal from the Bi matrix.

The stoichiometry of the As single crystal has been confirmed through elemental analysis using EDAX measurements. Figure 2(c) shows the observed EDAX spectra obtained from a mechanically cleaved As single crystal. The inset image displays the SEM image of the same crystal, while the inset table provides the atomic and weight percentages. The mechanically cleaved As single crystal is free from Bi or any other impurities, as the atomic percentage is determined to be approximately 99.87%, which falls within the margin of uncertainty. The inset image in Fig. 2(c) is the SEM image, which reveals the microstructural morphology of the As single crystal.

Furthermore, the SAED pattern of the synthesized As crystal has been recorded and is displayed in the upper image of Fig. 3(a). This image shows bright spots, confirming the crystalline structure of the synthesized crystal. Notably, the spots corresponding to (0 0 18) and (0 0 21) have been clearly marked in the image. The lower (0 0 3n) planes have not been marked, as their corresponding spots are obscured by the central bright spot. The bottom image of Fig. 3(a) presents the HRTEM image captured at1 nm resolution, which demonstrates the stacking of (0 0 3) planes. The inter-planar spacing for the (0 0 3) plane has been calculated to be 3.21Å, which closely matches the value derived from the XRD pattern.

To study the vibrational modes, present in gray-As, Raman spectroscopy has been performed at room temperature. Fig. 3(b) displays the observed results, showing two Raman peaks at 196.2cm$^{-1}$ and 255.74cm$^{-1}$, which are in accordance with the previous report [37]. This shows that the studied mechanically separated gray-As is free of embedded Bi impurity. DFT calculations are also performed to verify the vibrational modes observed theoretically. The phonon calculation was performed with norm-conserving pseudo-potential of PBESol type. The output of DFT calculation shows four characteristic frequencies for the D_3d(-3m) point group, out of which two modes are Raman active. These normal modes of oscillations are at frequencies 211.9cm$^{-1}$ and 253.5cm$^{-1}$ identified as doubly degenerate $E_g$ and $A_{1g}$ modes. The calculated normal mode frequencies are in line with the experimentally obtained characteristic frequencies. The inset shows the schematics of two normal modes of vibrations present in rhombohedral As.

Next, the single crystal of As is characterized for its electrical and magneto-transport properties. Fig. 4(a) shows the ρ-T plot at zero magnetic fields in the temperature range of 2 K-300K. The fitting with $\rho_0+AT^n$ in the low-temperature region (<75K) results in n=2.017 exhibiting quadratic behavior, highlighting the dominant electron-electron interactions in the Fermi Liquid regime. At higher temperatures (>100K), resistivity shows a linear behavior, exhibiting the presence of electron-phonon interactions. The result shows the metallic behavior of As crystal with e-e interactions being dominant over a wide range of temperatures up to 75K. Also, the residual resistivity ratio (RRR) given by $\rho_{300K}/\rho_{2K}$ is ~180, which is quite large. This validates the quality and the high metallicity of the synthesized crystal. Next, we study the ρ-T characteristics of As crystal under an applied transverse magnetic field. The inset of Fig. 4(a) shows the schematics of the geometry of current and applied field direction.

Fig. 4(b) shows the variation in resistivity with temperature in the range 2K-300K at applied transverse magnetic fields 0, 6, 10, and 14Tesla. The resistivity of As single crystal first decreases with decreasing temperature up to $T_{min}$, and then it starts increasing exhibiting the semiconducting behavior. This metal-to-insulator (MIT) transition is well known for similar semimetals under the applied transverse magnetic field [22,38-41]. The MIT transition temperature $T_{min}$ is obtained by the temperature derivative of resistivity. $T_{min}$ is the temperature at which dρ/dT change sign, above $T_{min}$ the derivative is positive (metallic), and below it is negative (insulator). $T_{min}$ increases as the ρ-T is measured at higher fields. The inset of Fig. 4(b) shows the log scale plot of ρ-T. The plateau regions in the ρ-T plot for the low-temperature range (<15K) under applied transverse magnetic field signifies the 2D nature of electron gas [39-41], supported by the magneto-transport measurements. The semiconducting behavior of the ρ-T plot under applied transverse field follows the relation $\rho(T)\propto \exp(E_g/(k_BT))$, where $E_g$ is the activation energy and $k_B$ is the Boltzmann constant. Using this relation, the activation energy induced by the applied magnetic field is calculated. Fig. 4(c) shows the ln(ρ) vs. 1/T plot. The slope is determined by fitting the linear part of the plot. The calculated activation energies from the obtained slopes are 1.52meV, 3.33meV, and 4.0meV at 6, 10 and 14Tesla, respectively. The magnetic activation, an upturn, and plateau in resistivity at low temperatures can be explained through the destruction of surface states in TIs due to the breaking of time-reversal symmetry as the magnetic field is turned on. In the TIs, the resistivity activation comes from an insulating bulk and the plateau from conducting surface states [40-42].

Fig. 5(a) shows the applied transverse field dependent resistivity ρ-H at 2K, 10K, and 20K of the As single crystal. Resistivity increases with the increasing applied field, showing the classical behavior of electrons under the Lorentz force. At low temperatures, 2K and 10K, and high fields, small oscillating humps are observable marked by arrows, and at 20K these humps do vanish. These humps in resistivity are known as SdH oscillations exhibiting the quantum 2D electron gas nature [42]. The SdH oscillations are extracted from ρ-H data by plotting the dρ/dH vs. 1/H as shown in the inset of Fig 5(a).

At low temperatures, as indicated by the plateau region, charge carriers behave as 2D gas. In the presence of a transverse magnetic field, 2D electron gas forms Landau Levels (LLs). The LLs split and get separated as the field is increased. The $n^{th}$ index of LL at magnetic field H implies that the conduction of charge carriers is through the $n^{th}$ LL. When Fermi energy lies in the middle of an LL, there are maximum charge carriers available for the conduction resulting in a minimum in resistivity. When the Fermi level lies in between the two gapped LLs, minimum charge carriers are available exhibiting maxima in resistivity. At a given field H, $n^{th}$ LL crosses the Fermi Level. In the infinite field limit, all LLs cross the Fermi Level. Therefore, as we increase the magnetic field the oscillation period becomes larger, and the involved $n^{th}$ LL is inversely proportional to the applied transverse magnetic field. So, in the infinite field Limit, the LL index must be zero. Since we have not gone to the very high magnetic field limit, the LL index starts with index 16. The landau levels are indexed by assigning integers to the maxima of observed oscillations in resistivity. Each maximum is separated by 1. In between the two maxima, a minimum is assigned integer +1/2. Following the above, the LLs are extracted from the inset of Fig. 5(a) and plotted as a function of the inverse of the transverse magnetic field, as shown in Fig. 5(b). All these LLs indices are linearly fitted through the Lifshitz-Onsagar quantization rule, which is given as

$$\frac{A_F \hbar}{eH} = 2\pi(n + \gamma) \qquad (1)$$

where $\gamma = {}^{1}/_{2} + \beta$, and $2\pi\beta$ is the Berry phase, $A_F$ is the cross-section area of the Fermi surface, e and $\hbar$ are electron charge, and Planck constant h divided by $2\pi$, respectively. Since the Berry phase is calculated from the LL fan diagram, to increase the accuracy and better linear fit, half-integer minima are also plotted and considered for fitting. The linear fitting yields a non-trivial Berry phase $\approx \pi$, which elucidates the presence of topological electrons with linear dispersion of the Dirac cone. These results are in accordance with the previous report [22]. The same non-trivial π Berry phase extracted from quantum oscillations has been observed in graphene [43] and (TIs) [44]. The linear fitting yields a slope of 218.37Tesla, which is the SdH oscillation frequency, and using equation 1, the $A_F$ is obtained as 208.2×$10^{16}$m$^{-2}$. The 2D charge carrier density is given by $n_{2D} = A_F/4\pi^2$ which yields $n_{2D}$ ≈3.49×$10^{11}$ cm$^{-2}$. The results obtained are close to that of a TI [45,46].

Though de Haas van Alphen(dHvA) and SdH frequencies are observed for different physical quantities, they have the same origin – the cyclotron orbit motion of electrons mapping the fermi surface of the material. So, the dHvA frequencies are computed by implementing DFT computed Fermi surface to SKEAF package[47]. Fig. 5(c) depicts the computed dHvA frequencies for the Fermi surface corresponding to the band number 10 as discussed later. Multiple frequencies are observable as the angle of applied magnetic field with c-axis varies. The observed SdH frequency of $B_\perp$=218Tesla has 2D character since the computed angle dependent frequency follow $B_\perp/\cos(\theta)$ for low angles (<20$^o$) as observed in other TIs $Bi_2Se_3$[48] and $Bi_2Te_3$[49], NdBi[50]. This elucidates the conduction of charge carriers through surface states present in As. For higher angles, a clear deviation from the 2D character evident, exhibiting the presence of bulk channels for the charge conduction, indicating dormant surface state conduction. This upholds the observed anisotropies in the measured SdH frequencies which is due to the distinct topography of the fermi surface. In comparison to pure TIs [48, 49, 51] which have single Fermi surface, As has multiple oscillation frequencies and new frequency branch appear as the magnetic field angle increases with the c-axis.

The As crystal exhibits large MR which is given as $\frac{R(H)-R(0)}{R(0)}$. At a given temperature, the MR% is calculated by multiplying MR by a factor of 100. Fig. 5(d) shows the MR% vs H plot in the applied transverse field range 0-14Tesla. At 2K, the observed MR% is of the order of $10^4$, which is higher than the MR% exhibited by rhombohedral Bismuth [52]. The observed MR% is one order less than the previous report [22] yet it is comparable to the MR in other large MR materials [1,39-41]. The MR does not saturate up to 14Tesla and above 9Tesla exhibits SdH quantum oscillations. The MR% is lower for higher temperatures and at 100K, the MR almost dies out in comparison to 2K.

Further, to understand the topological character of the rhombohedral Arsenic the first principle calculations have been performed. For the electronic band structure and projected density of states (PDOS) calculations, unit cell parameters and atomic positions are taken from Rietveld refined structure obtained from XRD data. Fig. 6(a) shows the PDOS of As with SOC parameters. They highlight that near the Fermi level, DOS mainly originates from the s and p orbitals. Notably, a large dip is observable in the DOS near the Fermi level, yet the non-zero value affirms the experimentally observed semi-metallic behavior. Extending our study, the electronic band structure has also been calculated along the extended high symmetric point path Γ→T →H2|H0→L →Γ→S0|S2→F2 → Γ in the first Brillouin zone (FBZ). The FBZ marked with the high symmetric point path is shown in Fig. 6(b). Along this extended path, the calculated band structure without(w/o) and with(w) SOC parameters are shown in Fig. 6(c) and 6(d), respectively. The orbitals projections s, $p_x$, $p_y$, and $p_z$ are also depicted in the plot (see supplementary info). At point L, a Dirac cone (not sharp) type-I is observable at ~0.5eV below the Fermi level. For the w/o SOC case, the direct band gap between bands 10 and 11 is found to be 65.7meV, whereas, with the SOC parameter, the same is found to be 72meV. In the extended path T →H2, another possible type-I Dirac cone is observable, but it is just above the Fermi level. Here, for w/o SOC bands, the Dirac cone is gapless, whereas the SOC lifts the degeneracy and induces a band gap of 180meV. The w/o SOC cone has s and all three-p orbitals characters. However, when the SOC parameter is included, the upper part of the cone has s and $p_x/p_y$ orbitals character dominant, whereas the lower part of the cone has s and $p_z$ orbitals character dominant. Thus, a change in orbital character due to SOC is an indication of band inversion, elucidating the topological nature of As, which is further substantiated by the non-trivial Z2 invariants and the surface state calculations. Two bands cross the Fermi level and corresponding to those, the calculated Fermi surfaces are depicted in Fig. 6(e) as plotted in Xcrysden. The yellow color surface is of hole type, whereas the teal color surface is of electron type as computed using the SKEAF package. Only one Fermi surface crosses the FBZ boundary, which is the main contributing band for electronic transport.

Fig. 6(f) depicts the total six phonon bands of rhombohedral As in the reciprocal space. In the close vicinity of the center of the Brillouin zone, the three bands show linear behavior identified as acoustic modes. At the Γ point, there are three optical modes also present, one is doubly degenerate identified as $E_g$ mode having a frequency of 185.0cm$^{-1}$ and the other optical mode is identified as $A_{1g}$ mode, having a frequency of 242.9cm$^{-1}$.

The Bloch states computed using density functional methods are wannierized in WANNIER90 software, where the Bloch wavefunctions are transformed from k-space to real space [28, 29]. The maximally localized wannier functions (MLWFs) are obtained with the disentanglement procedure having a tolerance of $10^{-15}$a.u. and the Wannier bands are computed with convergence cutoff $10^{-11}$a.u. Out of 36 bands, 16 spin-orbit coupled Bloch bands are wannierized with projections s and p orbitals and the band structure is reproduced in the range ±2eV about the Fermi level on the same size of k-mesh 8×8×8. This process also confirms that the bands near the Fermi level have the same orbital contribution as obtained in PDOS.

The computed MLWFs are implemented in Wannier Tools by constructing a model tight-binding Hamiltonian. Using the tight-binding model, the Band dispersion in the plane H0→L →Γ is computed on a dense k-mesh of 101×101 and the same is shown in Fig. 7(a). Fig.7(b) shows the dispersion energy gaps for bands 10 and 11. An inverted cusp-type dispersion centered at L high symmetric point shows that the energy dispersion is not linear but rather parabolic, which implies the presence of heavy relativistic fermions in Arsenic. This suggests that the system can be described by Dirac Hamiltonian which needs further analysis and verification by experimental measurement such as ARPES.

As the orbital projected band structure suggests a possible non-trivial surface state, the surface state spectrum has been computed applying the iterative Green's function method. A slab system of 40 Arsenic cells is constructed along the c-axis by setting the surface card to the (001) plane and the calculations are performed by taking 101 slices of one reciprocal vector. The 2D FBZ for the constructed slab system is shown in Fig. 6(b) along with the slab k-path. The calculated surface state spectrum for the top surface is shown in Fig. 7(c). A possible Dirac cone is observable along path D-Γ just above the Fermi level, and the edge state at the top surface of the slab is the red line emerging from the Dirac cone. A recent ARPES measurement on As [53] exhibits the surface states 200meV below the Fermi level at the $\bar{\Gamma}$ (here marked as A) point which is visible in our surface state spectrum simulation as shown in Fig.7(c). Fig. 7(d) depicts the slab bands calculated along the same path, which shows the presence of surface states. The blue and red curve is the surface state at the top and bottom of the Arsenic slab system respectively. The slab bands confirm the observed surface state at point A below the Fermi level.

Further to characterize the topology present in the rhombohedral As, the characteristic invariant Z2 is calculated. The rhombohedral As has a centrosymmetric space group. The band structure calculation shows it preserves the time-reversal symmetry as all bands computed with SOC parameters are doubly degenerate. Also, for the SOC bands the highest valence band and lowest conduction band are well separated. The topology in such systems that preserve time-reversal symmetry, is characterized by Z2 invariant [54]. Thus, the Z2 invariant is a suitable parameter for Arsenic, which is calculated using the parity of bands at Time reversal invariant momenta points, Fukui- Hatsugai method [55, 56]. Here, we have used Wilson's loop method to compute the Z2 invariant [19,57] where the Wannier charge center (WCC) evolved in FBZ. Based on the windings of WCC Z2 invariant is calculated. If there is an odd number of exchanges, a non-trivial (Z2=1) is assigned and for an even number of exchanges, a trivial (Z2=0) value is assigned. Four indices ($\upsilon_0$; $\upsilon_1$ $\upsilon_2$ $\upsilon_3$) are used to characterize the topological insulators given as

$$\nu_0 = (Z2(k_i = 0) + Z2(k_i = 0.5)) mod\ 2$$

$$\nu_i = Z2(k_i = 0.5\ plane)$$

where i=x,y,z. The first index indicates the strong topology whereas the rest indicates the weak topology present in the system. From the above method, the Z2 index for As is (1,111). This dictates that As is a strong topological system. Our result is consistent with the Z2 invariant previously reported by the band parity method at TRIM points [23,24]. This not only supports the previous results but also confirms the accuracy of the WCC method. The topological states are regarded as robust against external perturbation [58]. To put this under test for As surface states, further the addition of disorder and impurity to As lattice system simulation are suggested.

## Conclusion

Summarily, the single crystalline gray-As samples are synthesized by the Bismuth flux method. The obtained single crystals are well characterized by XRD and TEM/SAED for their phase purity. EDAX shows that the synthesized As crystals are free from any foreign impurity. The vibrational modes are studied by Raman spectra and the two Raman active modes are identified by DFT calculations. Transport measurements reveal the Fermi liquid behavior of electrons at low temperatures which shows 2D nature under applied transverse magnetic field. The magneto-transport measurement shows giant MR, with SdH oscillations at low temperatures and high fields. The Landau Level fan diagram is calculated and fitted with the Lifshitz-Onsagar relation resulting in a non-trivial $\pi$ Berry phase. This shows the presence of the topological nature of charge carriers in the As crystals with linear dispersion Dirac cone. The theoretical calculation of dHvA frequencies gives insight into the conduction by surface states which agrees well with the observed SdH oscillation frequencies. The non-trivial topology present in gray-As crystal is well supported by the orbital projected band structure calculation and shows band inversion due to SOC present in rhombohedral As. The surface state spectrum shows the presence of edge states present at both, the top and the bottom surfaces of the Arsenic slab system. The non-triviality of these edge states is shown by computing the Z2 invariant. Since As preserves TRS, the Z2 invariant is calculated using the first principal methods. The Z2 invariant index is (1,111), which implies As is a strong TI.

**Table-1**

Parameters obtained from Rietveld refinement:

| Cell Parameters | Refinement Parameters | DFT Unit Cell |
|---|---|---|
| Cell type: Rhombohedral<br>Space group: R-3m (166)<br>Lattice parameters:<br>a= b=3.761(6)Å<br>& c= 10.542(3) Å<br>$\alpha=\beta=90^o$, $\gamma=120°$<br>Cell volume: 129.154 $Å^3$<br>Density: 8.203 $g/cm^3$<br>Atomic coordinates:<br>As1 (0 0 0.22849) | $\chi^2=2.22$<br>$R_p=7.2$<br>$R_{wp}=10.0$<br>$R_{exp}=6.71$ | Cell vectors (Ang):<br>1.880796 -1.085878 3.514063<br>0.000000 2.171757 3.514063<br>-1.880796 -1.085878 3.514063<br>ATOM_POSITIONS frac<br>As 0.22850 0.22850 0.22850<br>As 0.77150 0.77150 0.77150 |

**Figure Captions**

**Figure 1: (a)** and **(b)** Two-step heat treatment diagram followed for the synthesis of As crystals

**Figure 2: (a)** XRD pattern taken on a thin crystal flake of synthesized As single crystal. Inset shows the As crystals cleaved from Bi (As) matrix **(b)** Rietveld refinement of PXRD pattern of As. The inset shows the unit cell generated by VESTA. **(c)** EDAX spectra of synthesized As single crystal, in which the inset depicts FESEM image of surface morphology of As single crystal.

**Figure 3:(a)** The upper image shows the SAED pattern of As crystal and the bottom one shows the stacking of (0 0 3) planes. **(b)** Room temperature Raman spectra of As crystal with schematic vibrational modes shown in the inset.

**Figure 4: (a)**Temperature-dependent resistivity of As crystal at zero field. **(b)** ρ-T plot at different transverse fields showing MIT transition. **(c)** ln(ρ) vs 1/T plot linearly fitted at the different applied transverse fields.

**Figure 5: (a)** The magneto-transport of gray-As at 2K, 10K, and 20K. The inset shows the SdH oscillations extracted at 2K and 10K. **(b)** Landau Level fan diagram: LL index (n) vs. 1/H plot. **(c)** The DFT computed dHvA frequencies for the lower band crossing the Fermi level **(d)** MR% of As crystal at different temperatures.

**Figure 6:** **(a)** Orbital projected DOS of bulk rhombohedral As. **(b)** The k-path in the FBZ is guided by the green arrow. **(c)** and **(d)** The orbital projected band structure of rhombohedral As for w/oSOC and wSOC parameters, respectively. **(e)** The Fermi surface plot in the FBZ depicts electron and hole pockets. **(f)** The calculated phonon spectrum of rhombohedral As.

**Figure 7:** **(a)** The inverted cusp type band dispersion about L point **(b)** Dispersion band gap plot corresponding to computed band dispersion. **(c)** Surface state spectrum computed of Arsenic for the slab system constructed along 001 plane. **(d)** Slab bands show the presence of surface states near the Fermi level.

**Acknowledgment**

The authors would like to thank the Director of the National Physical Laboratory, New Delhi for his keen interest. The authors acknowledge Dr. J. Tawale and Ms. Sweta for SEM/EDAX and Raman spectroscopy measurements, respectively. N. K. Karn and Kapil Kumar would like to thank CSIR, India for the research fellowship and the Academy of Scientific and Industrial Research, Ghaziabad for Ph.D. registration. This work is supported by in-house project numbers OLP 240832 and OLP 240232. PARAM Rudra, a national supercomputing facility, at Inter-University Accelerator Centre (IUAC), New Delhi, has been used to obtain the results presented in this paper. The authors acknowledge Md. Zubair for the fruitful discussion regarding Z2 invariant calculation.

**CRediT authorship contribution statement**

**N.K. Karn:** Writing – original draft, Methodology, Investigation, Software, Calculation, Formal analysis, Data curation. **Kapil Kumar:** Synthesis, Writing – review & editing, Methodology, Data curation. **Geet Awana:** Writing – review & editing, Methodology, Data curation. **Kunal Yadav:** Writing – review & editing, Methodology, Data curation. **S. Patnaik:** Supervision, Writing – review & editing. **V.P.S. Awana:** Writing – review & editing, Supervision, Project administration, Methodology, Conceptualization.

**Data Availability Statement:**

All the data associated with the MS will be made available on reasonable request.

**Conflict of Interest Statement and Declaration:**

The authors have no conflict of interest. The authors declare that they have no known competing financial interests or personal relationships that could have appeared to influence the work reported in this paper.

**Supplementary information:** Not Applicable

**Ethical approval:** Not Applicable

Figure 1(a)

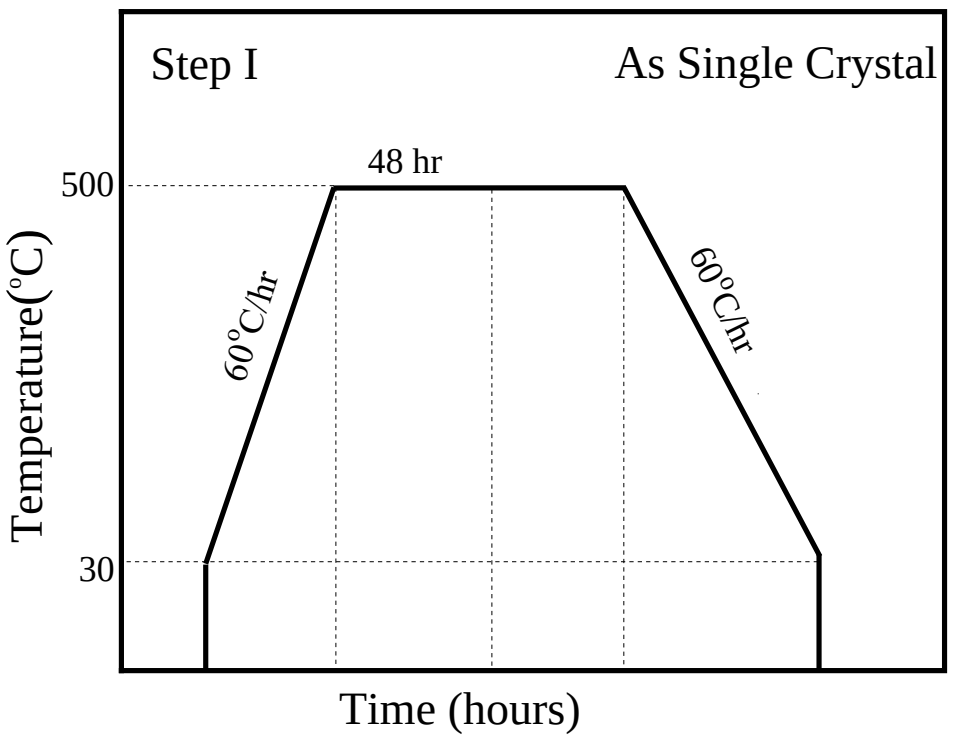

(b)

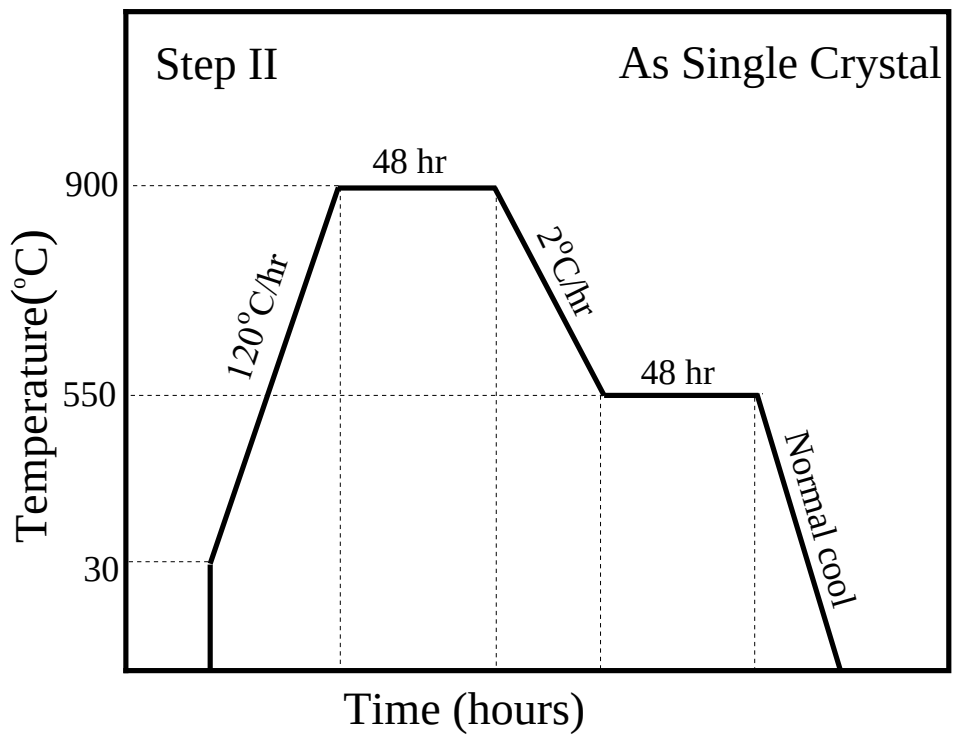

Figure 2 (a)

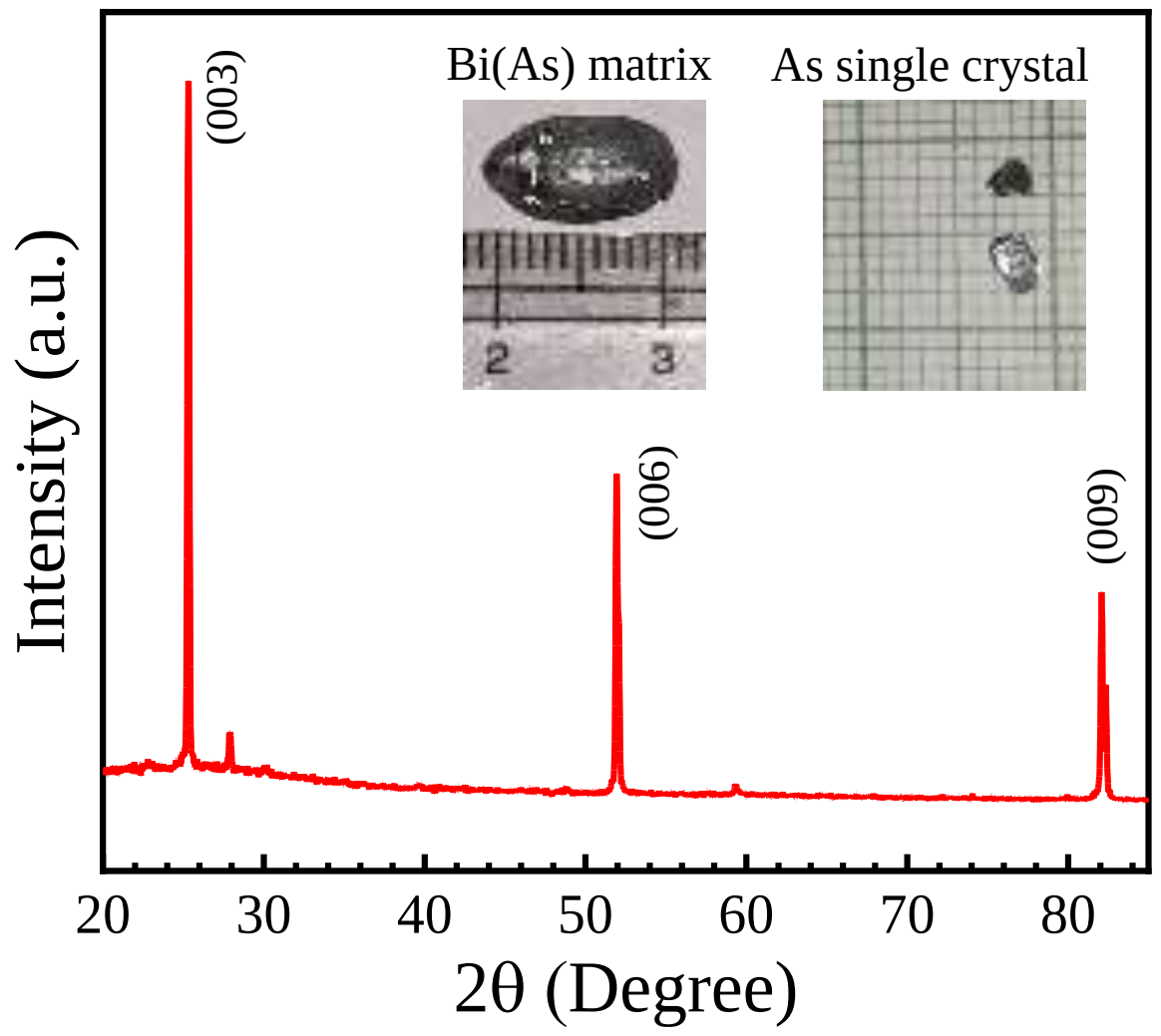

(b)

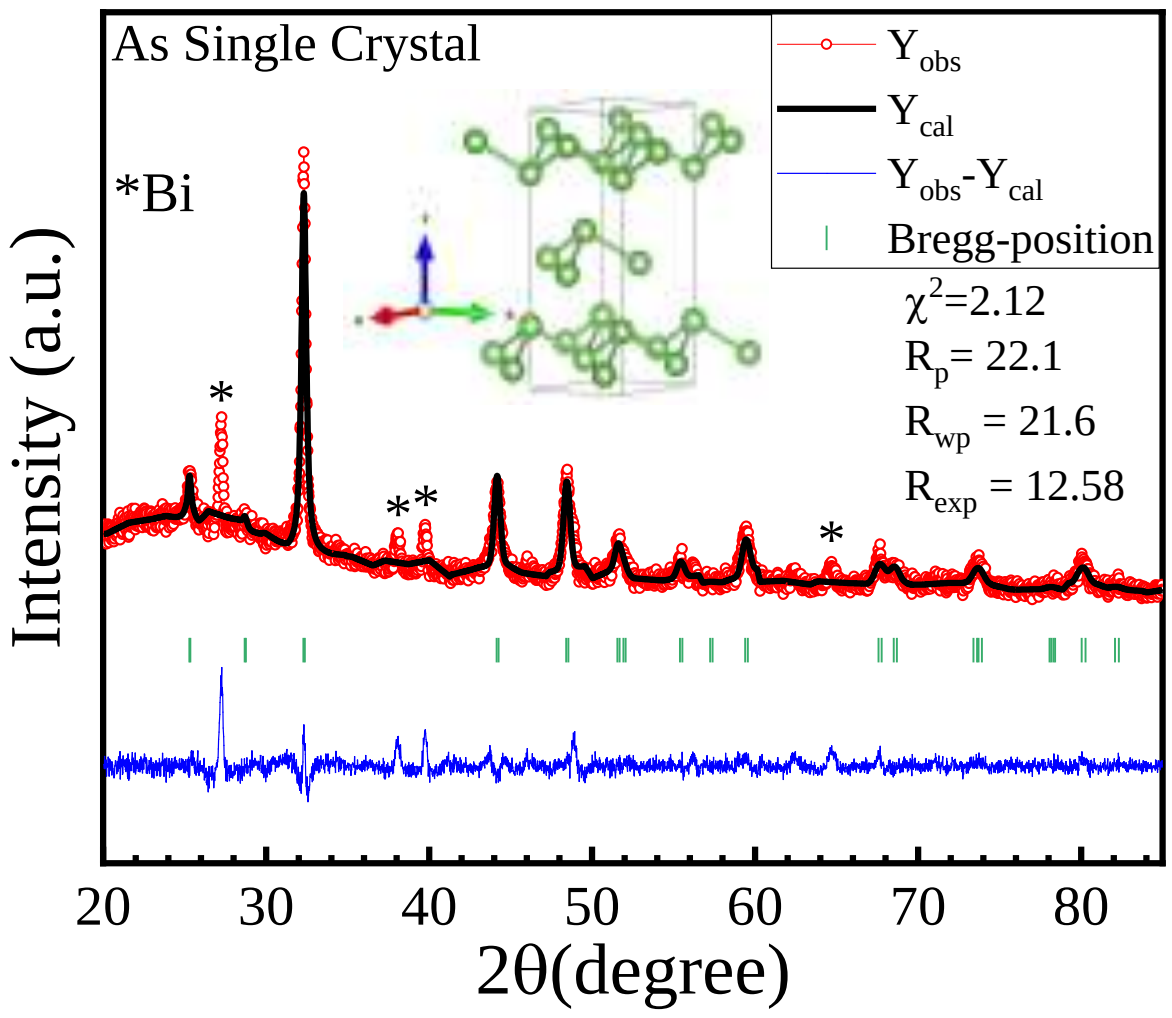

(c)

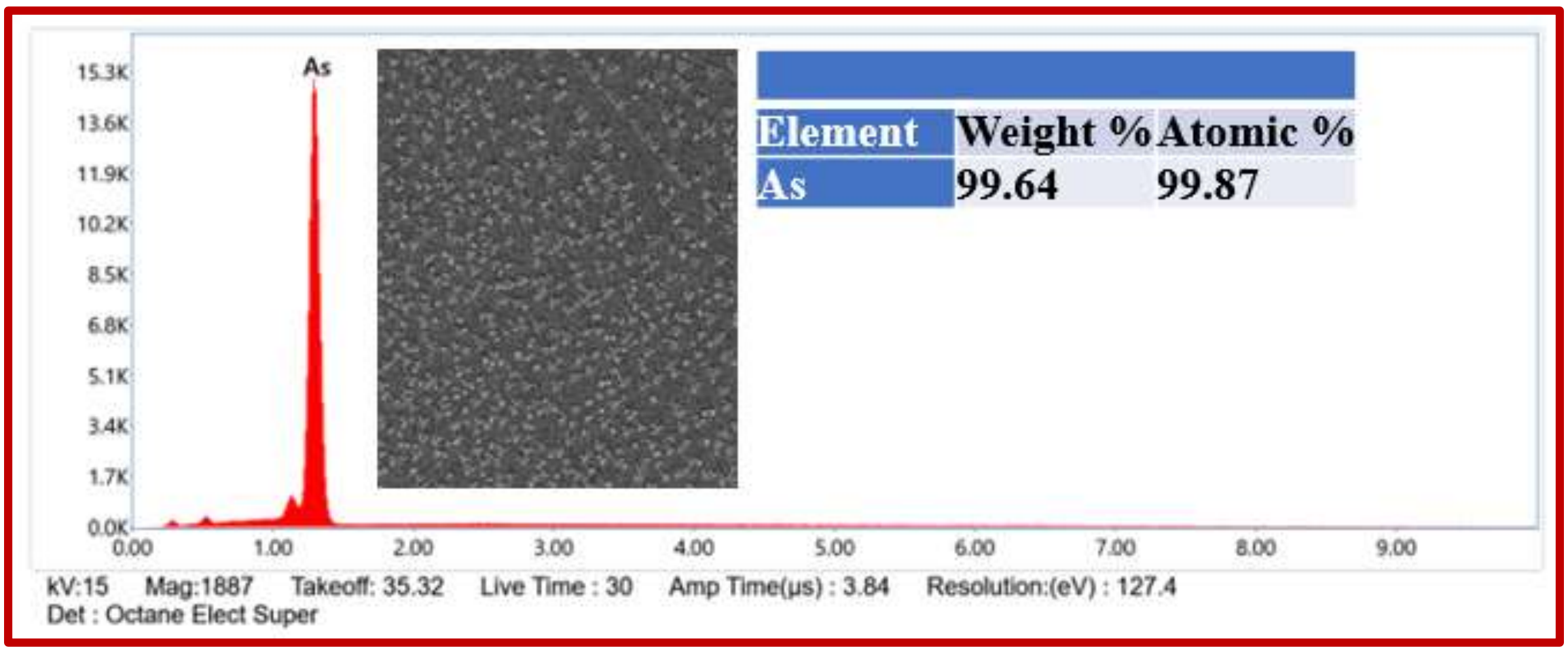

Figure 3(a)

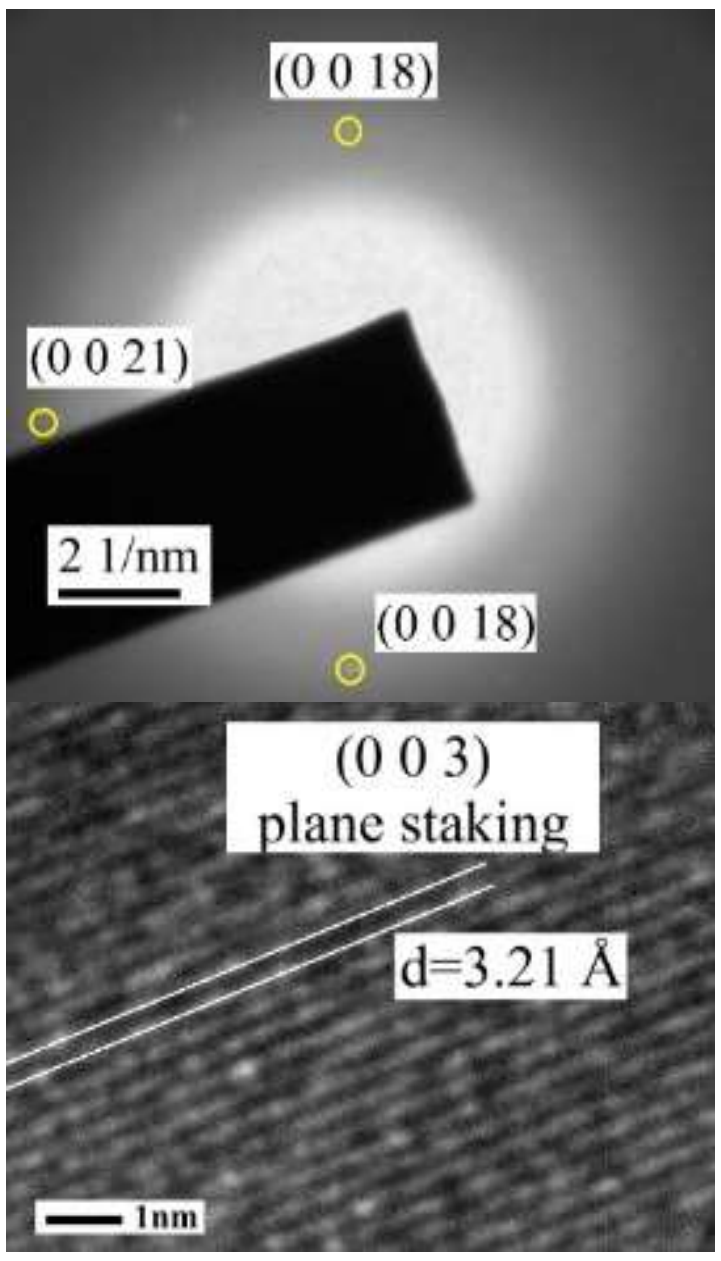

(b)

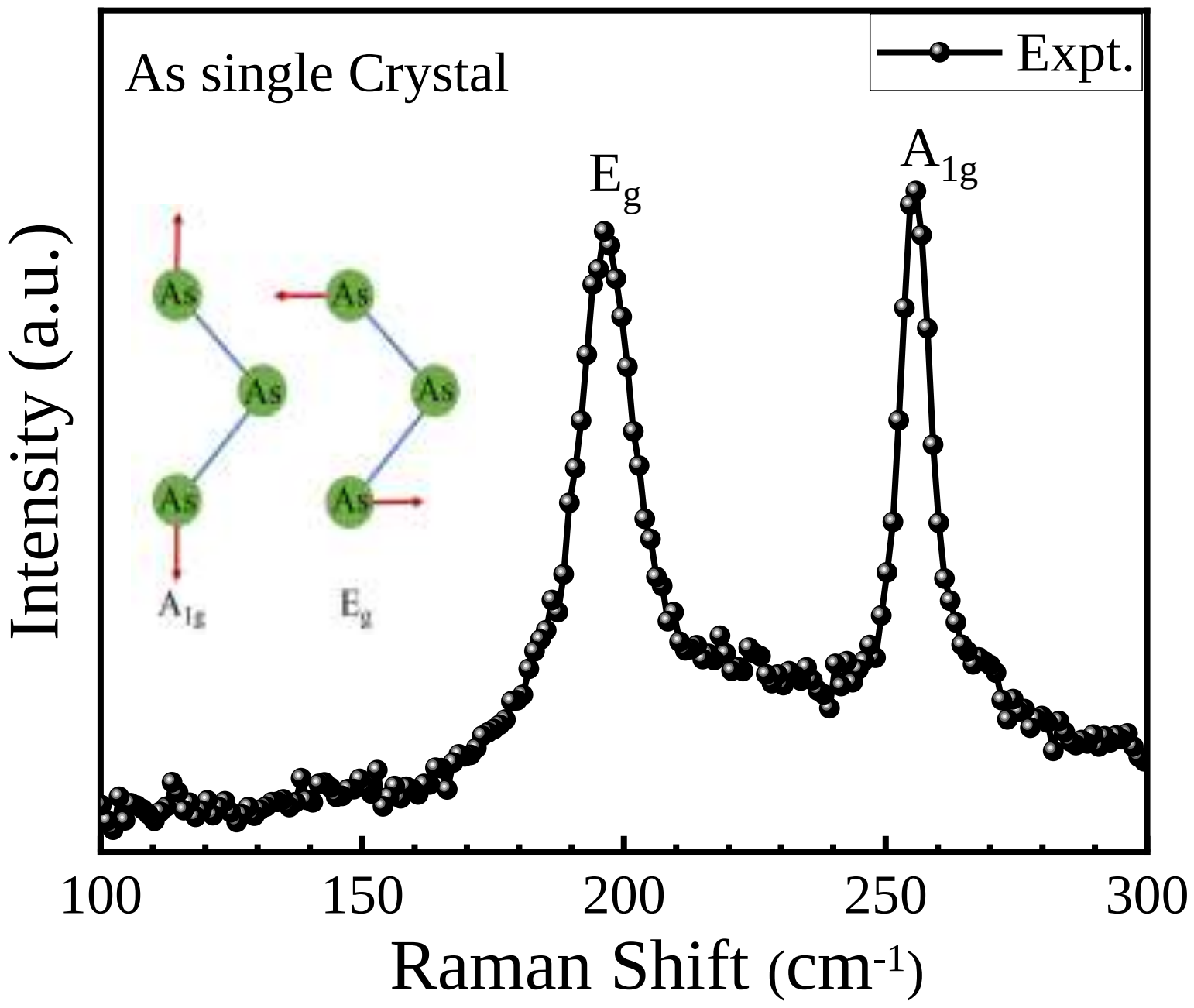

Figure 4 (a)

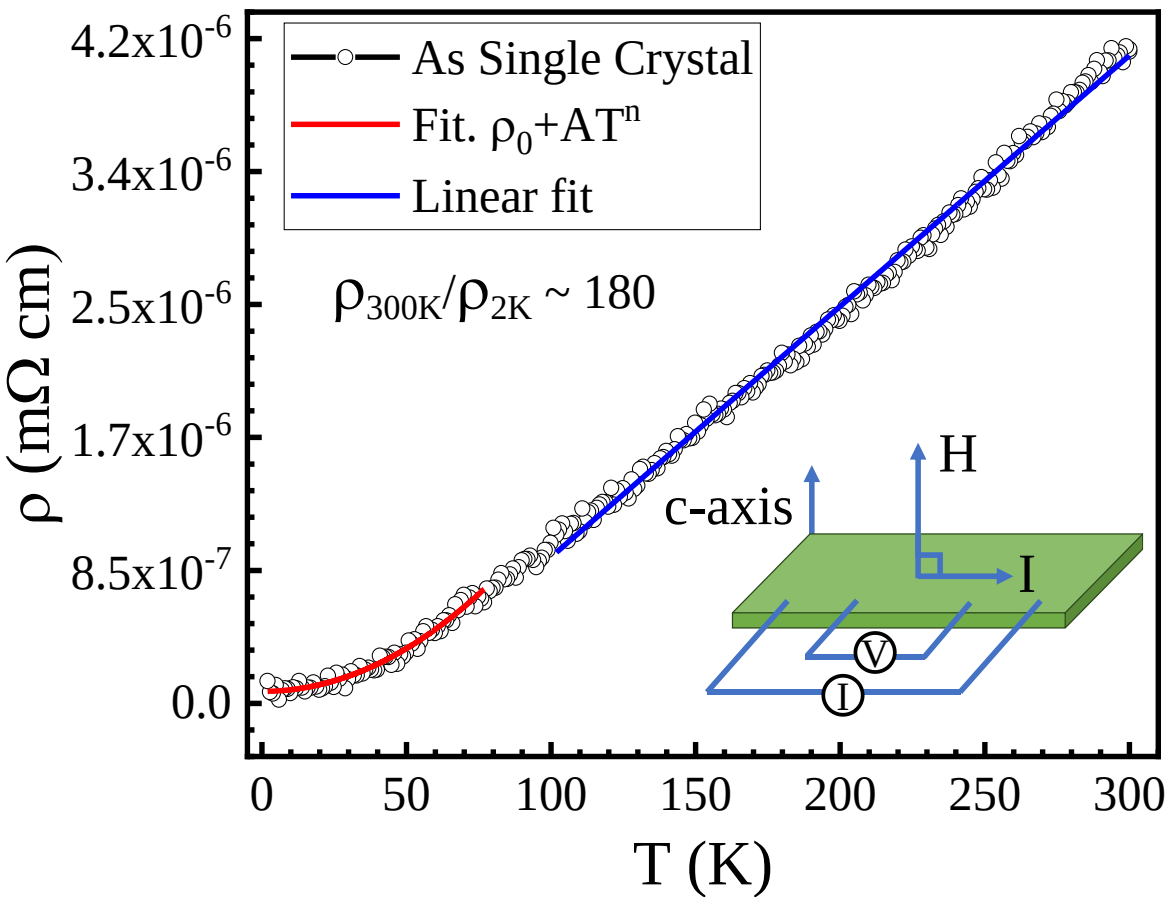

(b)

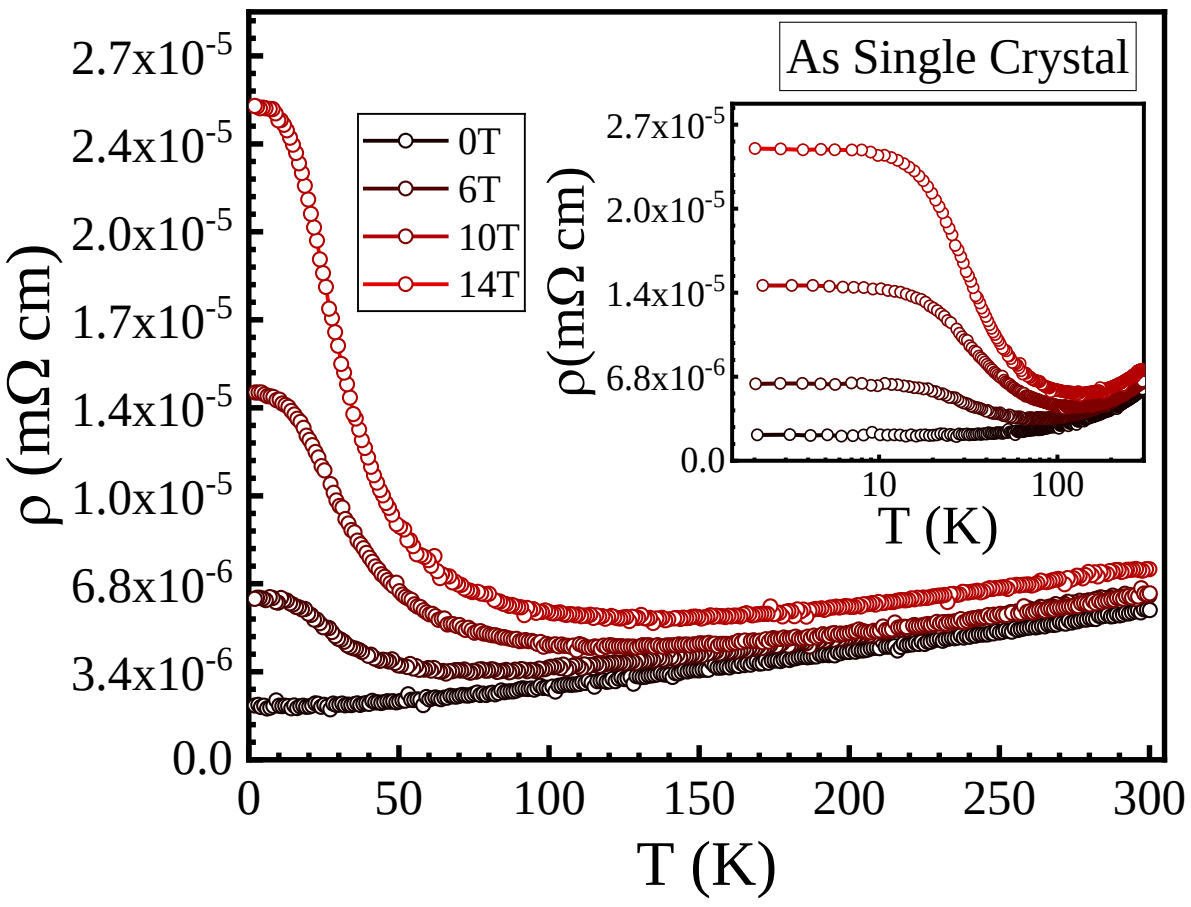

(c)

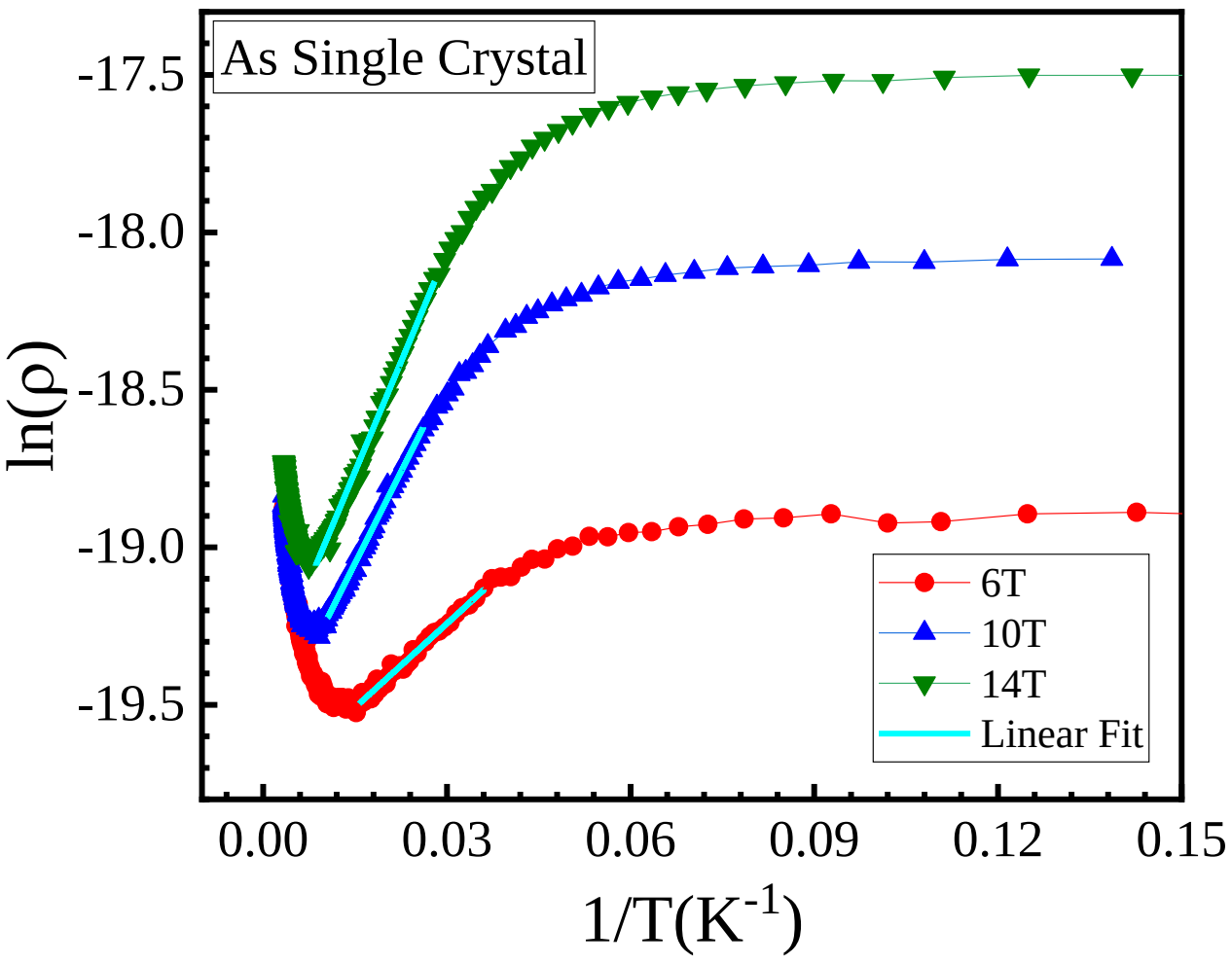

Figure 5(a)

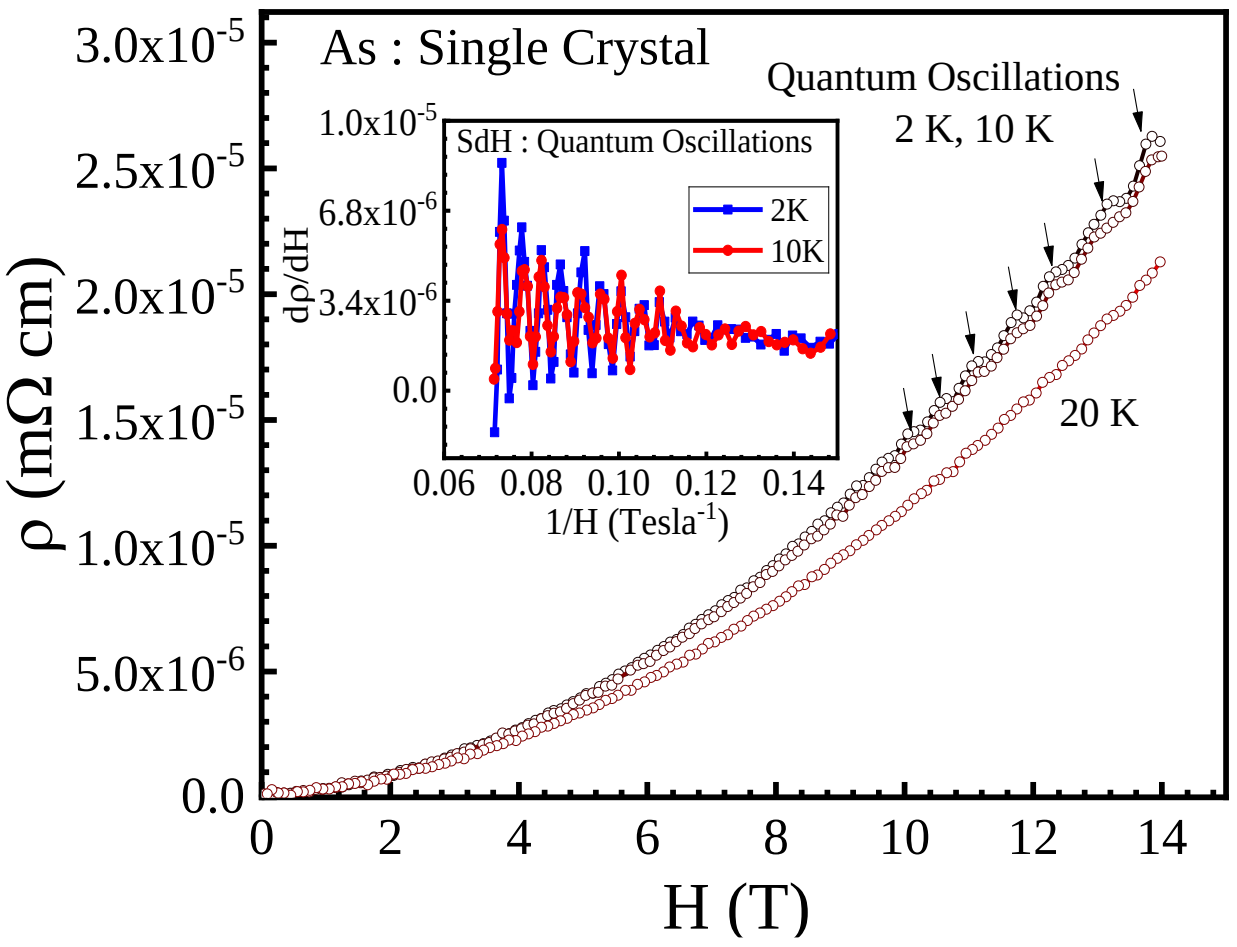

(b)

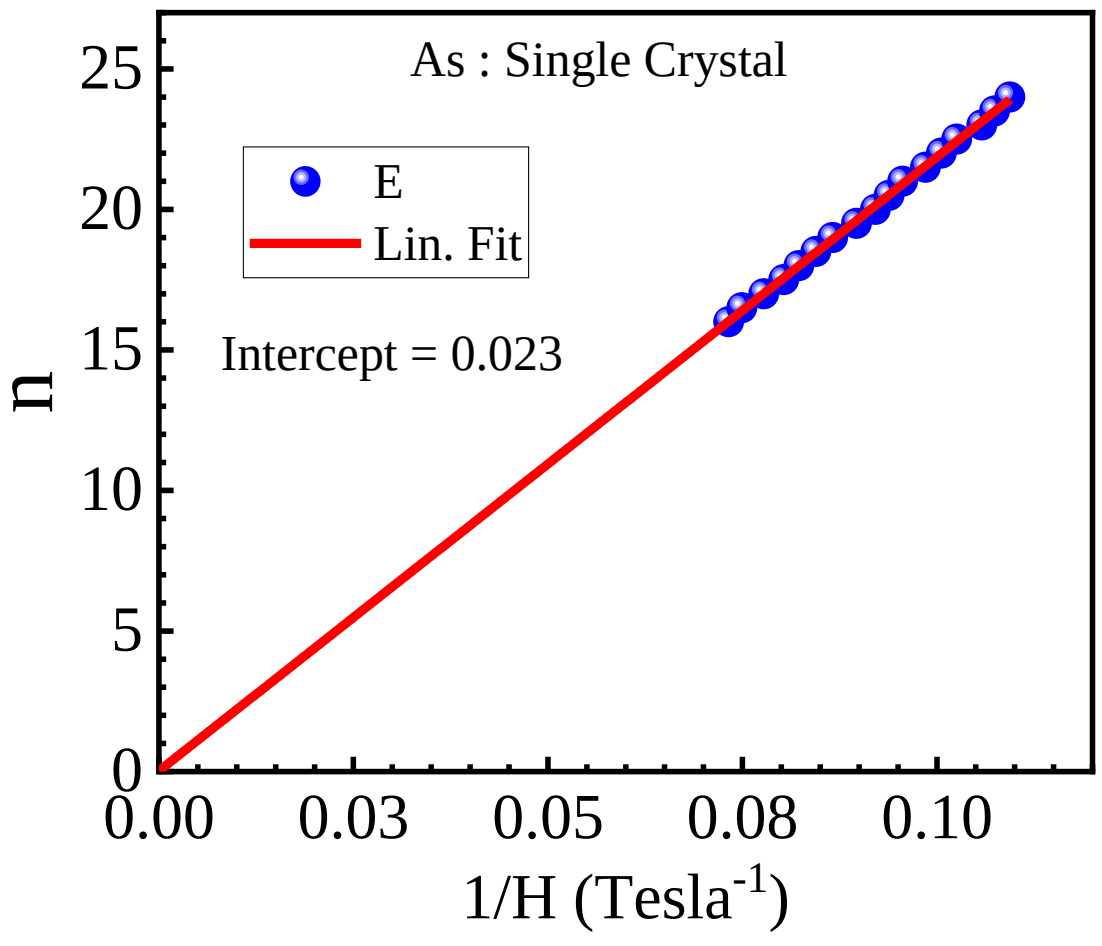

(c)

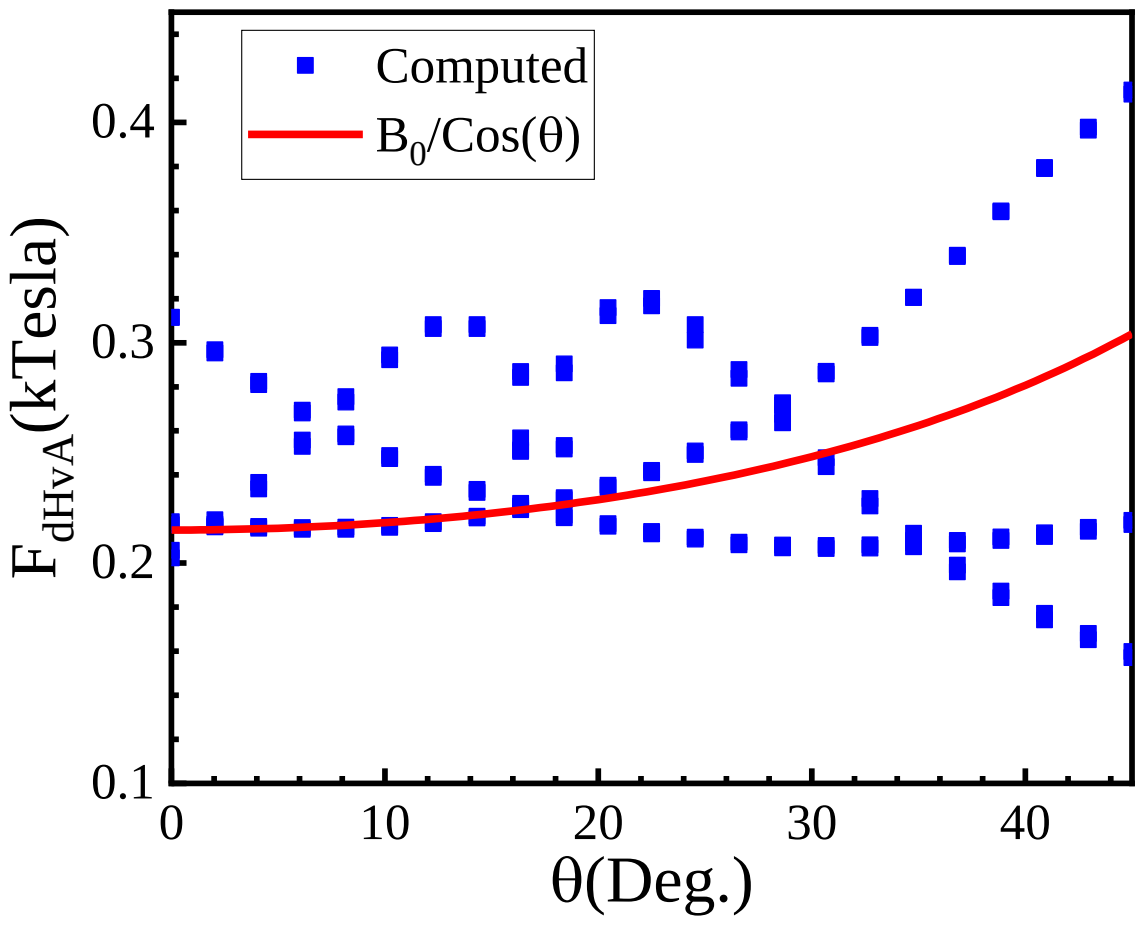

(d)

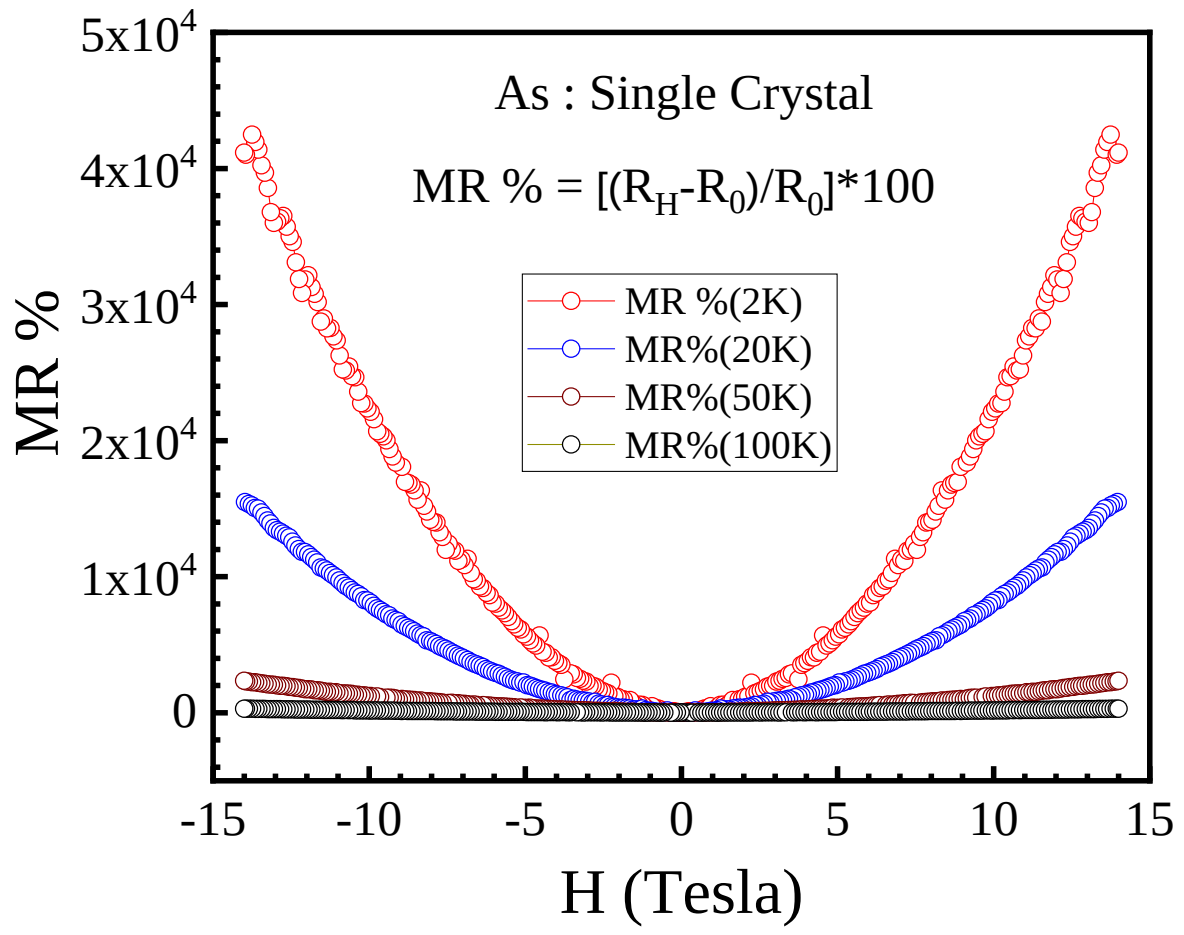

Figure. 6

(a)

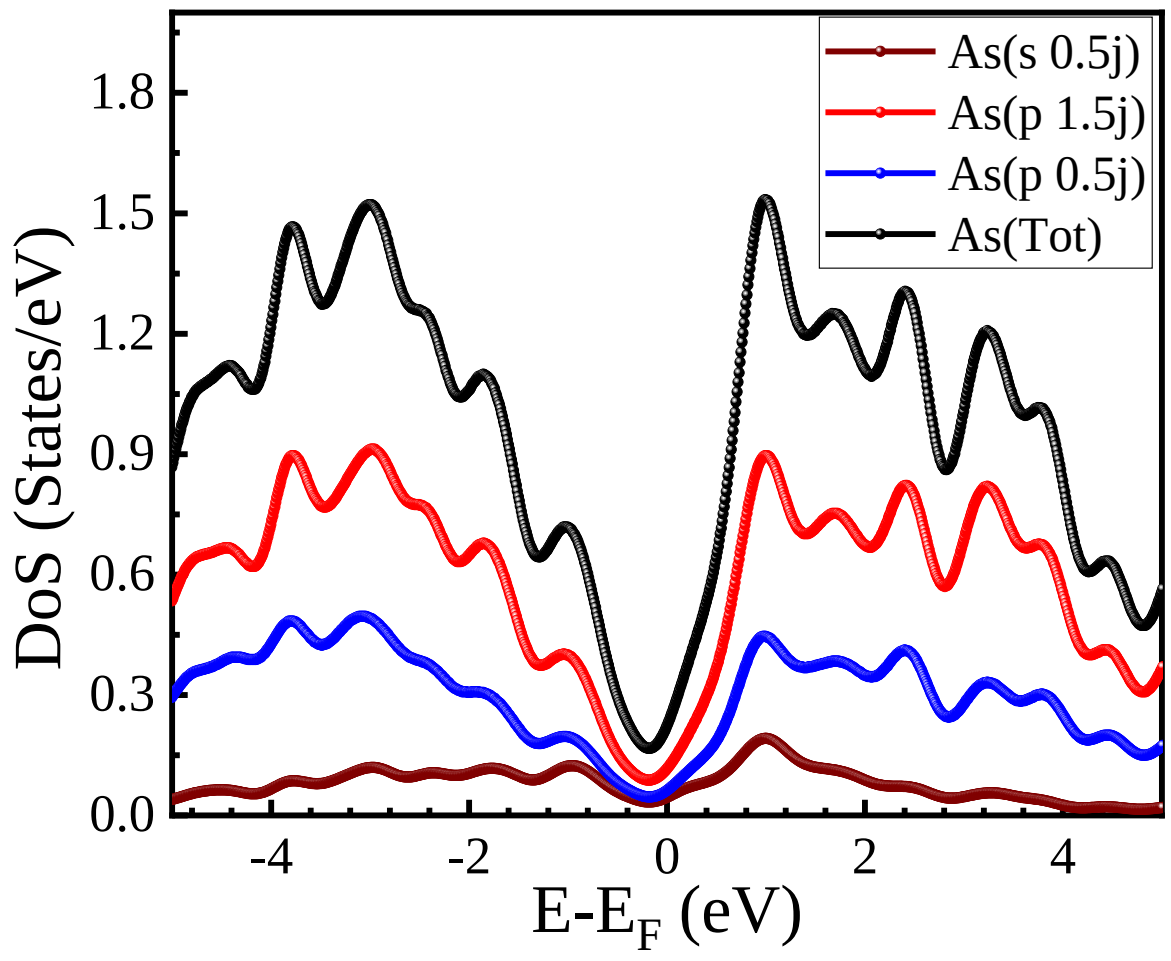

(b)

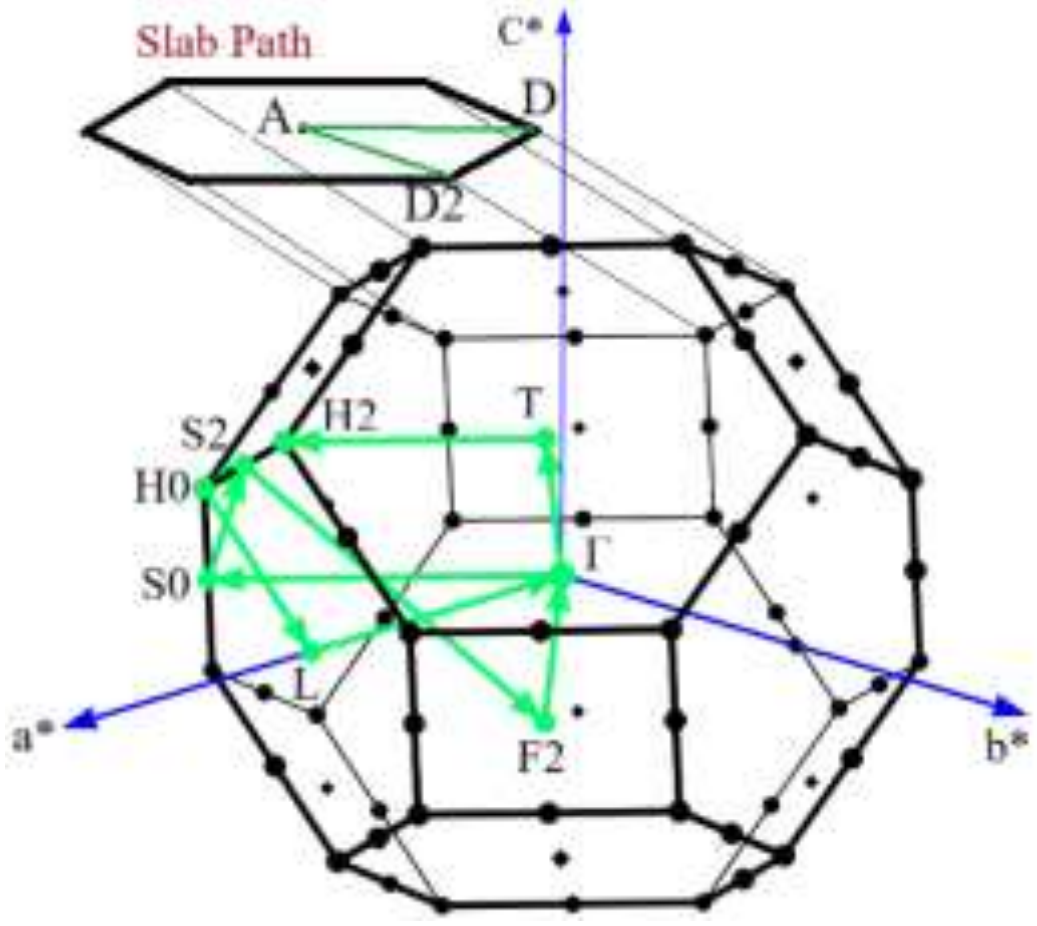

(c)

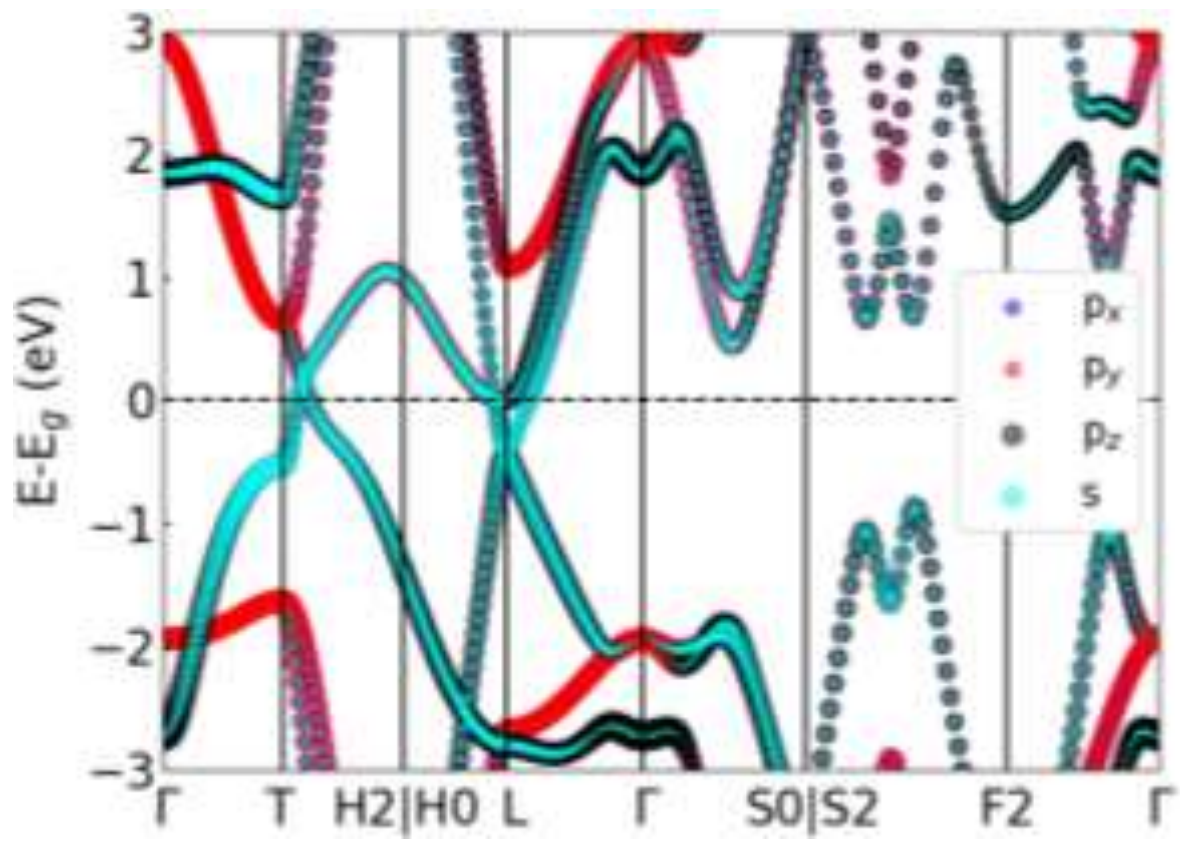

(d)

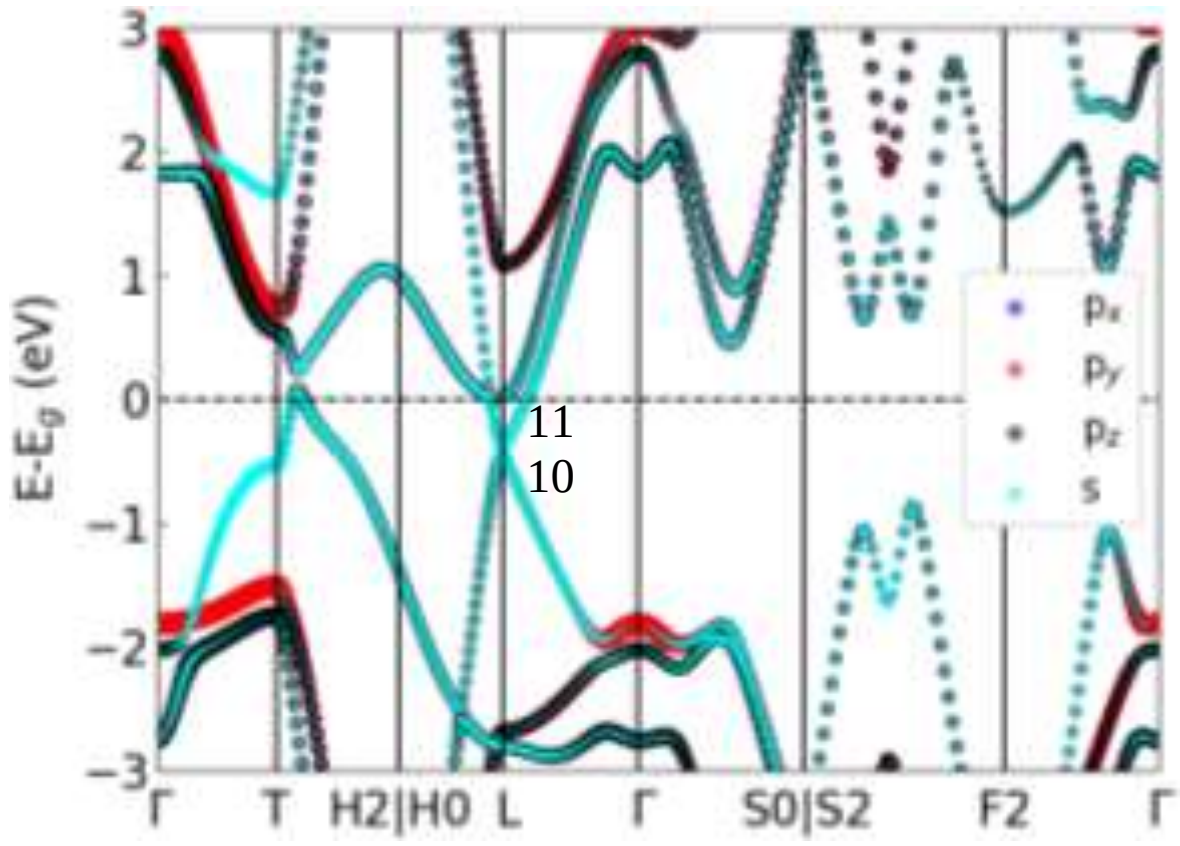

(e)

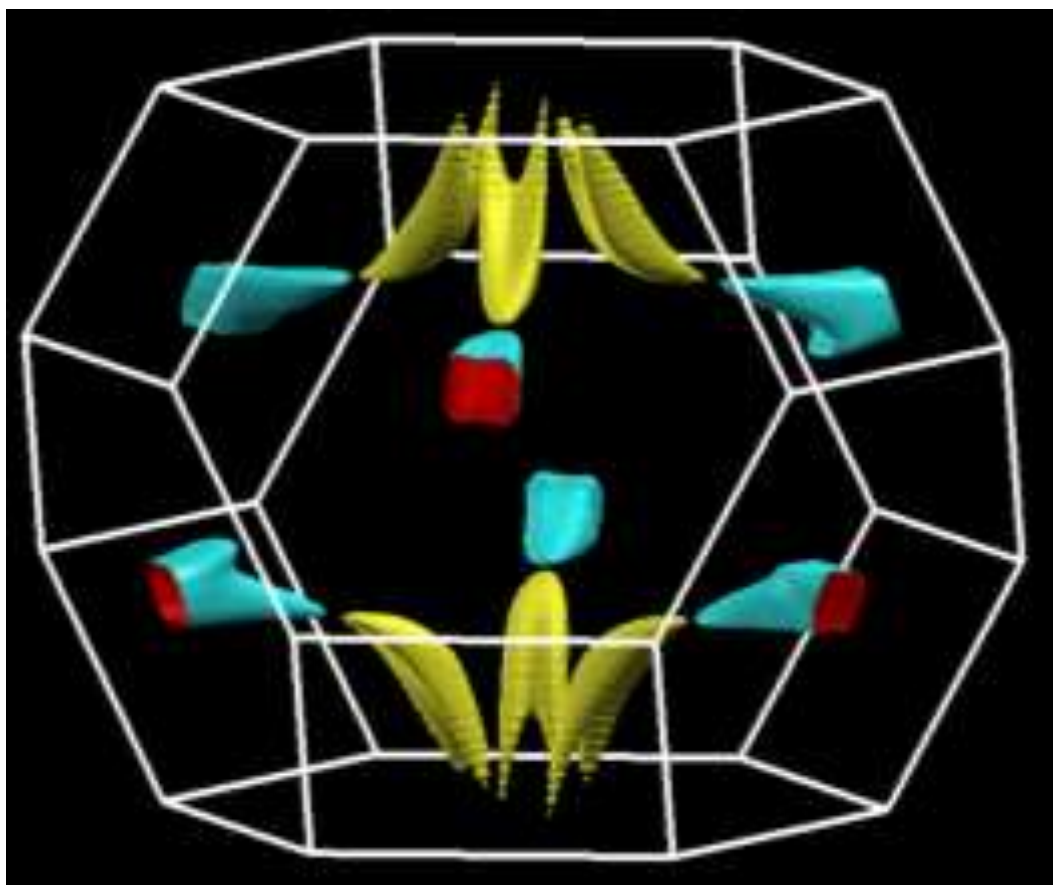

(f)

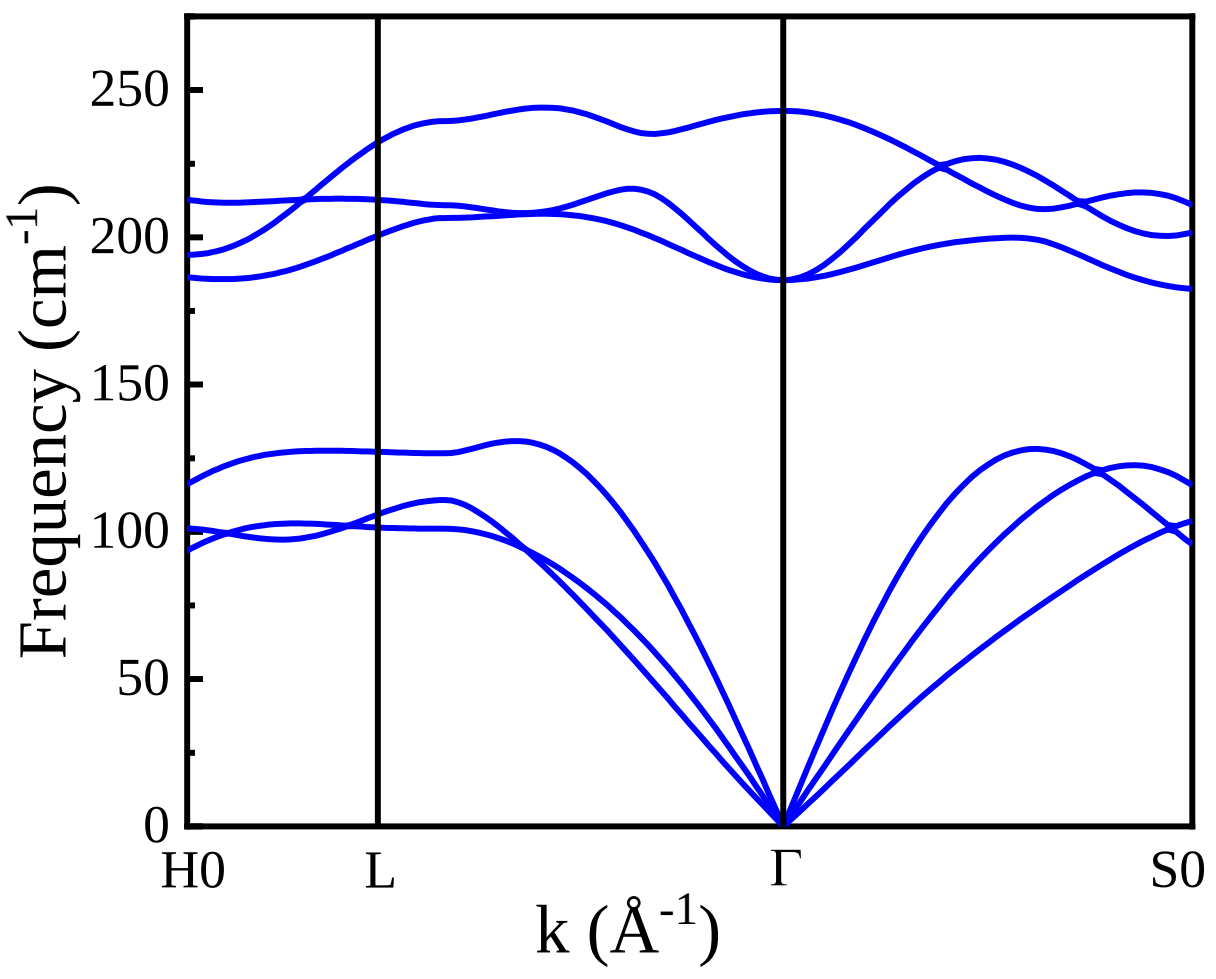

Figure. 7

(a)

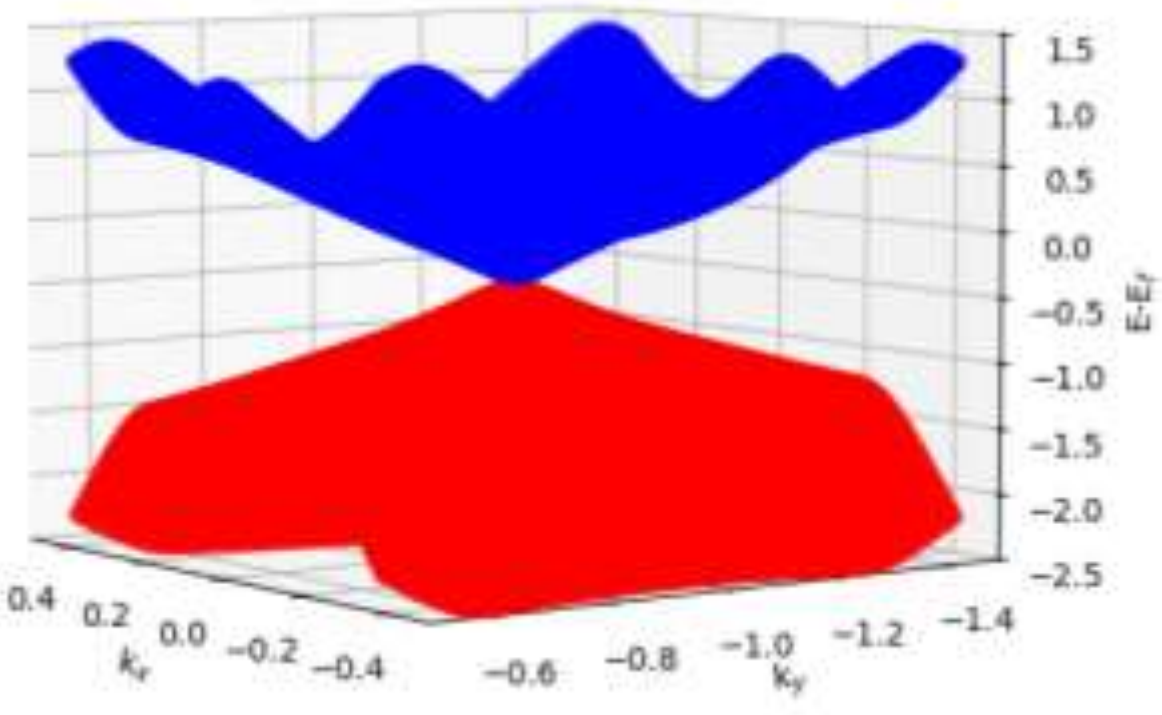

(b)

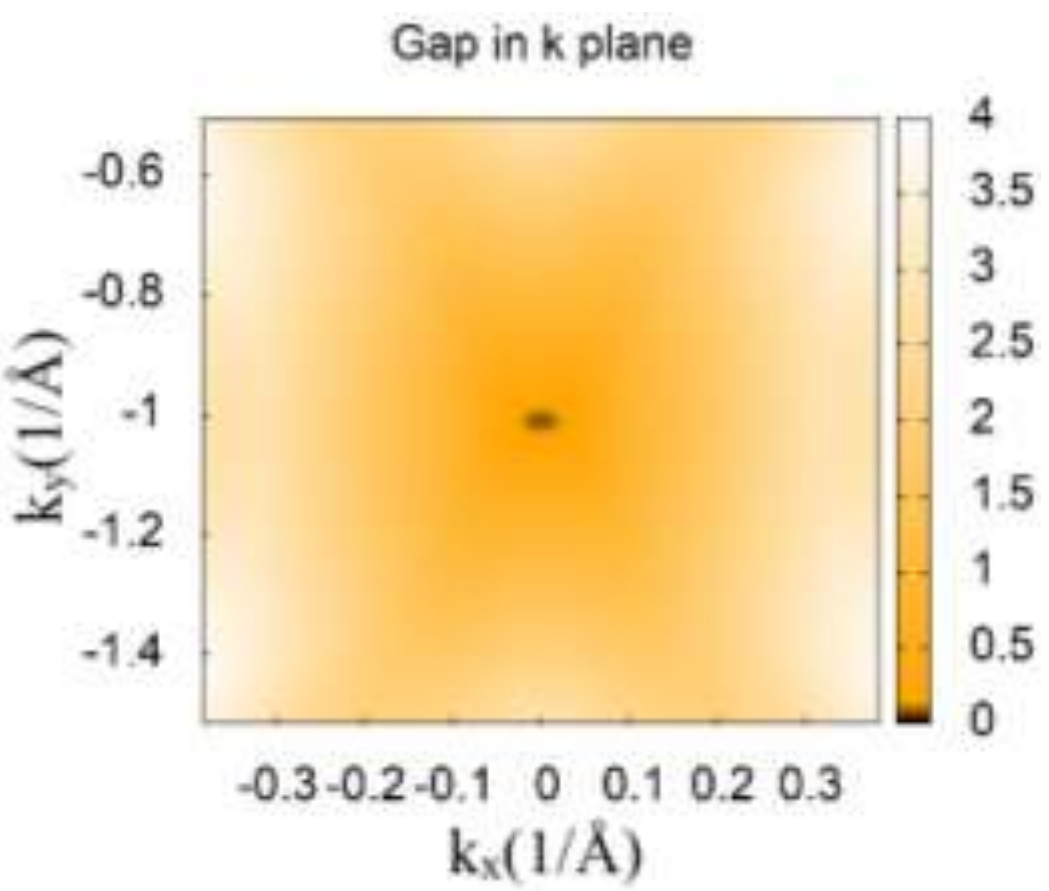

(c) (d)

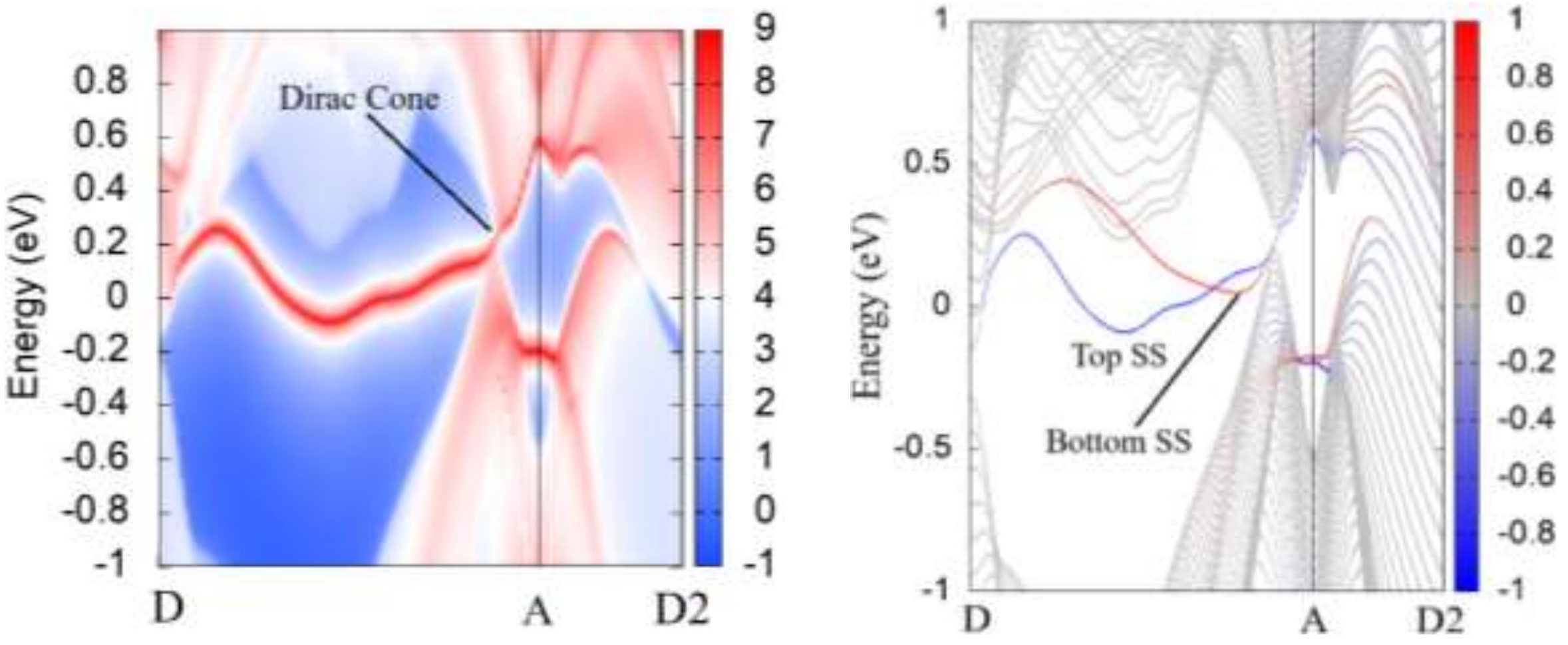